\documentstyle[12pt,aasms4]{article}

\begin{document}

\title{Testing An Identification Algorithm for Extragalactic
OB Associations Using a Galactic Sample}
\author{Christine D. Wilson and Karen J. Bakker}
\affil{Department of Physics and Astronomy, McMaster University,
Hamilton Ontario Canada L8S 4M1}

\begin{abstract}

We have used a Galactic sample of OB stars and associations to test 
the performance of  an automatic grouping algorithm designed to identify
extragalactic OB associations.
The algorithm identifies the known Galactic OB
associations correctly when the search radius (78 pc) is defined by the
observed stellar surface density, which suggests that
the sample of Galactic OB associations constitutes a reasonably
uniformly-identified sample. 
Roughly 25\% of the groups identified automatically which
contain more than 10 stars actually comprise two or more unrelated
OB associations, which raises the concern that the largest extragalactic
associations identified via this algorithm 
may also represent multiple associations. 
Galactic OB associations identified with a 78 pc search radius have
diameters that are $\sim$3 times larger than OB associations identified with
a 22 pc search radius in M33. Applying the smaller search
radius to the Galactic data matches both the sizes and the
number of member stars between the two galaxies quite well. Thus,
we argue that this and similar algorithms should
be used with a constant {\it physical} search radius, rather than
one which varies with the stellar surface density. Such an approach would
allow the identification of differences in the giant molecular cloud
populations and star formation efficiency under most circumstances.

\end{abstract}


\section{Introduction}

Determining the properties of OB associations in galaxies is important
for understanding how massive stars form. Differences in the properties
of OB associations from one galaxy to another may indicate differences
in the sizes of the molecular clouds from which the associations
form, in the star formation efficiency in the molecular clouds, or 
in the relative numbers of high and low mass stars that are formed.
However, obtaining a reliable comparison of
association properties from one galaxy to another requires that
the associations are identified in a consistent
and unbiased way. Early identifications of OB associations were
based on identifying clumps of blue stars by eye using photographic
plates (e.g. M33, \markcite{hs80}
Humphreys \& Sandage 1980). However, comparison
of the properties of OB associations identified in this way is
extremely difficult unless similar quality plates are available
for all galaxies and one person has done the identification
(\markcite{h86}Hodge 1986). 

Recently, two automated techniques for identifying
OB associations in galaxies have become available (\markcite{b91}Battinelli 
1991;
\markcite{w91}Wilson 1991). In addition, the availability of CCD data 
allows us to select stars based on strict color and magnitude
cutoffs for input to the algorithms. Applying such an automated
algorithm to CCD data for M33 and NGC 6822 showed that the
number of member stars and diameters were quite similar for OB associations
in the two galaxies (\markcite{w92}Wilson 1992), in contrast to earlier 
photographic results which suggested that the OB associations in
NGC 6822 were a factor of three larger than those in M33 
(\markcite{h77}Hodge 1977).
These automated algorithms represent a standardized,
objective approach to extragalactic OB association identification.
However, we have little guarantee that the associations thus identified are
true physical groupings of stars of similar age.
The algorithms cannot distinguish between true associations
and random projections of unrelated stars, although comparison of the
actual data with groups identified from a random distribution of stars
suggests that the larger associations are unlikely to be chance
projections (\markcite{w91}Wilson 1991). In addition,
we might expect that the algorithm would merge 
associations in close proximity to one another
into a single larger association. 

One way to test
the algorithm is to apply it to a sample of O stars and known OB 
associations in the solar neighborhood. Ideally, we would like to
use a large sample of O stars for which membership in OB associations
has been identified unambiguously using radial velocity and proper
motion data. Although several samples of O stars exist
(\markcite{he84}Humphreys
\& McElroy 1984, \markcite{bh89}Blaha \& Humphreys 1989,
\markcite{gcc82}Garmany et al.
1982), the identification of OB associations 
 is rather heterogeneous. While membership in associations
within 1 kpc of the Sun is usually confirmed using photometry,
radial velocity, and proper motion data (\markcite{g94}Garmany 1994),
membership in more distant associations is likely to be assigned based
only on position in the sky and a distance estimated from
spectroscopic parallax (\markcite{g94}Garmany 1994,
\markcite{h78}Humphreys 1978, \markcite{ca71}Conti \& Alschuler 1971).
An additional complication is that the boundaries of the OB
associations are generally somewhat arbitrarily defined, and in fact are
rectangular (\markcite{g94}Garmany 1994). Thus in some sense 
the identification of OB associations beyond about 1 kpc from the
Sun may be almost as ill-defined as ``by-eye'' identifications of
OB associations in external galaxies. However, the Galactic
data do have a greater
spatial dynamic range; at 3 kpc, one arcminute corresponds
to 1 pc, while at the distance of M33 one parsec corresponds
to 0.25$^{\prime\prime}$. 
We can therefore assume that these ``by-eye'' identifications are
more accurate than those in external galaxies.
Thus it is reasonable to attempt to use this
rather heterogeneous data set to test how well an automatic identification
algorithm for OB associations manages to reproduce the {\it a priori}
known Galactic OB associations.

In this paper we apply the automated identification algorithm of
\markcite{w91}Wilson (1991) to the combined OB star catalog
of \markcite{he84}Humphreys
\& McElroy (1984) and \markcite{bh89}Blaha \& Humphreys (1989), and
also to the more limited catalog of \markcite{gcc82}Garmany
et al. (1982) and compare the results with the OB associations
identified in the catalog.  We also apply the algorithm to a
random distribution of the stars as was done for the extragalactic
data (\markcite{w91}Wilson 1991). 
The catalog and data processing are described
in \S 2 and  the algorithm and 
results are presented in \S3. A comparison of the 
properties of the Galactic associations with those of M33 and NGC 6822
is presented in \S 4. The paper is summarized in \S 5.

\section{The Galactic Catalog}

One data set used to test the automated identification algorithm
are the catalog of OB stars and supergiants in associations, clusters,
and in the field from \markcite{he84}Humphreys
\& McElroy (1984) and \markcite{bh89}Blaha \& Humphreys (1989).
This catalog contains (among other information) the absolute magnitude,
distance, Galactic latitude and longitude, and association membership
(if available). 
We also used the catalog of Galactic OB stars by \markcite{gcc82}Garmany et al.
(1982), which is available in machine-readable format through
the Astronomical Data Center.
We limited our analysis to stars within 3 kpc of the Sun.
To simulate the type of data available for
external galaxies, we first projected the stars onto the plane of the
Galaxy. To match the data available for M33 (\markcite{w91}Wilson 1991),
we then applied a faint cutoff in absolute magnitude of $M_v = -4.4$.
Note that, given the large scatter in $M_v$ for a given O spectral
type (0.5-0.8 mag, \markcite{gs92}Garmany \& Stencel 1992), 
this magnitude cutoff will eliminate some main sequence stars across the
full range of O spectral types.

The resulting distribution of stars for the
combined Humphreys catalog is shown in Figure~\ref{fig-1}a.
One obvious feature in this diagram is the presence of many
arc-like features; these features are the OB associations, whose
member stars are assigned a common distance in the catalog.  The distribution
in the plane of the sky of the stars in a given OB association shows
that they are extended in both dimensions (\markcite{gs92}Garmany
\& Stencel 1992), and thus a reasonable
assumption is that the associations also have some depth in the
radial direction. Hence for each OB association we estimated its
diameter from the length of the arc in Figure~\ref{fig-1}a and
then randomly distributed the stars about a circular region with
the given diameter. The results are shown in Figure~\ref{fig-1}b and
are much more similar in appearance to extragalactic data
(cf. \markcite{w92}Wilson 1992).

\section{Re-Identifying Galactic OB Associations Using an Automatic Algorithm}

Candidate Galactic OB associations were identified using the ``friends of
friends'' algorithm (\markcite{w91}Wilson 1991). The mean surface density of
stars, $\Sigma$, was used to calculate the search radius, which is defined
as $R_s = \sqrt{1/\pi\Sigma}$. For each star, the star list
was searched to identify all stars lying within one search radius of
the initial star. Any new stars identified in this way were
checked in their turn for other stars lying within one search radius, until
no more stars were added to the group. For this subset of the catalog, there
are 1484 stars lying within 3 kpc of the Sun, so the search radius
used was 78 pc. 
Galactic OB associations identified using the algorithm
are compared with the previously
identified associations in Figures~\ref{fig-2}-\ref{fig-4}. 
The algorithm places all stars in groups, some of which contain only
one or two members. A minimum number of members ranging from 3 to 10
is commonly used when identifying significant OB associations in external
galaxies. \markcite{w91}Wilson (1991) used a minimum member cutoff of 10, based
on the sizes of groups identified from a random distribution of stars,
while \markcite{b91}Battinelli (1991) adopted a minimum member cutoff of 3, so
as not to miss the smallest associations.
We also include results for a minimum member cutoff of 6, which
was chosen based on
our estimate of the contamination by random groups (see below).
Figure~\ref{fig-2}
compares all previously known 
OB associations containing at least 3 stars brighter than
$M_v = -4.4$ mag with the associations containing at least
3 members identified by the automatic algorithm. 
Figure~\ref{fig-3} shows associations containing
at least 6 bright stars, and Figure~\ref{fig-4} shows associations
containing at least 10 bright stars. A comparison of the associations
identified by the algorithm with the actual data shows that only a single
association was missed by the algorithm in the case of associations
with 10 or more bright stars, and that this association is recovered
in the sample with 6 or more stars.
Thus the Galactic OB associations identified in
previous surveys  appear to constitute an internally consistent sample,
i.e. one that has been identified using a uniform scale length.

In some cases new associations are identified that
were not known {\it a priori}; these associations may in fact be
previously unidentified OB associations (due to their relatively
low stellar content, large extent in radial distance, and, in
some cases, proximity to the us), or they may
be chance associations of physically unrelated stars. 
Assuming the latter, the relative fraction
of ``false'' associations identified by the algorithm is
45\% for a 3 member cutoff, 25\% for a 6 member cutoff, and 10\% for
a 10 member cutoff. 
We can compare this result with the number
of associations identified from a random distribution of stars
with the same surface density and area. 
The relative number of associations identified 
from the random distribution as compared to the
actual Galactic data is 300\% for a 3 member cutoff, 40\% for a 6 member
cutoff, and 0\% for a 10 member cutoff. Thus the random distribution
of stars predicts significantly more associations with small numbers
of members than are seen in the actual data. This result can be
easily understood: the true stellar distribution is significantly
clumped (Figure~\ref{fig-1}), and, with large numbers of stars grouped
into several rich associations, there are many fewer stars to
populate the field region. Thus a truly random distribution of
stars may give an overly pessimistic estimate of the number of
false associations identified using this automatic algorithm.

The mean diameter of the OB associations identified
in previous surveys is about 150 pc (\markcite{gs92}Garmany
\& Stencel 1992). 
The median diameter of the associations with more than
3 members identified automatically is 100 pc, while
 the median diameter of the associations with more
than 10 members identified with the automatic algorithm
is  about twice as large (180 pc).
Roughly 20-25\% of
the Galactic associations identified by the algorithm
are actually two or more nearby associations that the algorithm cannot
physically separate. 
This result suggests that mis-identification of two nearby associations
as a single association may be a significant problem for the larger
associations. Such a mis-identification could have important
consequences for our understanding of massive star formation:
if two unrelated associations with different ages were mistakenly
identified as a single association, we might conclude that a 
single OB association had undergone more than one episode of massive
star formation. In addition, if an association had triggered the
formation of second association through the process of self-propagating
star formation, if the associations lie too close together in the sky,
the two associations might not be separated and thus
the age gradient might not be recognized. 

The results of the analysis of the catalog of \markcite{gcc82}Garmany 
et al. (1982) 
are qualitatively similar to the results discussed above. For this
catalog there are 
are 404 stars lying within 3 kpc of the Sun, so the search radius
used was 150 pc. This search radius did a reasonable job of identifying
the previously known OB associations. For this data set,
the relative fraction
of ``false'' associations identified by the algorithm is
25\% for a 3 member cutoff, 5\% for a 6 member cutoff, and 0\% for
a 10 member cutoff, while the relative number of associations identified 
from the random distribution 
is 135\% for a 3 member cutoff, 25\% for a 6 member
cutoff, and 0\% for a 10 member cutoff. 
The median diameter of the associations with more than
3 members identified automatically is 160 pc, in good agreement
with previous studies.
However,  the median diameter of the associations with more
than 10 members identified with the automatic algorithm
is  twice as large (360 pc). This large diameter is due
to the presence of four associations which are made up two or more
nearby associations and also the presence of four associations
which contain only a single association but have been extended
to include nearby field stars. Because of the small
number of associations with more than 10 members in this catalog,
the relative fraction of
associations that are in fact composed of multiple nearby
associations increases from $\sim 15\%$ for groups with at least
3 members to $\sim 50\%$ for groups with at least 10 members.

\section{Comparing Galactic and Extragalactic OB Associations} 

The main source of
subjectivity in the use of this automated algorithm 
 is the choice of the search radius.
\markcite{w92}Wilson (1992) 
identified OB associations in NGC 6822 using two different
search radii, the first determined from the surface density of blue
stars (39 pc), and the second being the same physical
radius (22 pc)
that was used in a similar study of M33 (\markcite{w91}Wilson 1991).
The main difference in the associations identified with the
two different search radii occurs in 
the mean diameter of the associations, which is
twice as large with the first (larger) search radius. However, the
search radii calculated from the stellar surface density are
fairly similar in the two regions studied, with that for NGC 6822 only 
1.7 times larger than that for M33.
In comparison, the search radius calculated for the Galactic data
set is $\sim 3.5$ times larger than the M33 search radius.

Unless the presence of massive stars acts as a trigger for subsequent
episodes of star formation (as may be the case in starburst or
very active galaxies), the overall surface density of massive stars
in a region probably has little bearing on the formation of OB
associations. Thus tying the search radius to properties of the
large-scale population, either through the stellar surface density
\markcite{w91}(Wilson 1991) or through maximizing the number of groups
\markcite{b91}(Battinelli 1991), is difficult to justify on physical 
grounds for normal galaxies.
To first order, let us assume that OB associations form a population that
is fairly homogeneous in its properties, particularly its size distribution,
from one galaxy to
another. Then it is more logical to identify OB associations 
using a search radius of a fixed {\it physical} length,
rather than one which is tied to the stellar surface density.
Since OB associations form in molecular clouds, 
this situation might be expected to hold if galaxies had similar
molecular cloud populations and if the star formation
efficiency in the individual molecular clouds was roughly
constant. 
While the universality of the molecular cloud 
population has yet to be tested in many galaxies, the giant
molecular cloud population in M33 does appear very similar to that
of the Milky Way (\markcite{ws90}Wilson \& Scoville 1990). Data on the
star formation efficiency of individual clouds is even rarer,
but some recent results (\markcite{wm95}Wilson \& Matthews 1995) 
suggest that the star formation efficiency
in the two brightest HII regions in M33 is not much higher than
the star formation efficiency in the Orion molecular cloud
\markcite{el91}(Evans \& Lada 1991). Thus the assumption that the 
populations of OB
associations in M33 and the Milky Way have similar
properties is justifiable, although such an assumption may not
be true for other galaxies. In particular, molecular
clouds in dwarf irregular
galaxies appear to be smaller on average than those in spiral galaxies
(\markcite{rlb93}Rubio et al. 1993).

In fact, the use of a search radius of fixed physical length does not
prevent us from obtaining information about average cloud sizes and
star formation efficiencies in galaxies. For example, consider two
galaxies which contain molecular clouds of the same size but which
have different star formation efficiencies. Identifying the associations
with a fixed search radius would produce associations of similar sizes in both
galaxies, but with different numbers of member stars, i.e. the galaxy with
the larger star formation efficiency would have associations with more members
(Figure~\ref{fig-5}). In comparison, if we study two galaxies with the same 
star formation efficiency but with different average cloud sizes, the
algorithm would find larger associations containing more members in
the galaxy with the most massive clouds. Since for giant molecular clouds the
average density scales inversely with the cloud radius \markcite{sss85}(Sanders
et al. 1985), the total mass of the cloud goes as the radius squared. 
Thus the surface densities of stars in the associations in the two galaxies
would be the same. Finally, consider two galaxies with different cloud 
sizes and different star formation efficiencies, with the galaxy containing
the largest clouds having a lower star formation efficiency. Only in this
situation are the results of the algorithm potentially unclear, since
the search radius may be too small to identify properly the large, low
surface density associations in the second galaxy. Thus by adopting a
fixed search radius for all galaxies, we can under many circumstances
investigate both the relative cloud sizes and star formation efficiencies
in different galaxies.

Under these assumptions, 
the choice of the search radius would in principle be entirely arbitrary.
To determine the effect of the search radius on the properties of
the identified associations, we have analyzed the Galactic
catalog, the M33 catalog, and the NGC 6822 catalog with search
radii of 12, 22, 41, 78, and 150 pc. Analysis of the Galactic
data set shows that the median association radius is approximately
linearly proportional to the search radius. For the M33 data set
the median association radius increases more steeply with the search
radius (roughly as the 1.6 power), while for the NGC 6822 data set
the increase may be slightly less than linear. For all three galaxies
the median number of stars per association is less sensitive
to the search radius, and increases only by a factor of 2-3 as the 
search radius is increased by a factor of seven. The median radius
and median number of stars in the associations agree quite well for all
three galaxies when search radii in the range of 12-22 pc are used. For larger
search radii the M33 associations are significantly larger than the Galactic 
associations. This result can probably be explained by the much higher stellar
surface density in the M33 region than in the solar neighbourhood, since
we would expect the radius of an association to
increase faster with increasing search radius 
for a region with a higher surface density of field stars.
We cannot probe search radii much below 12 pc, since
a search radius of 6 pc is only 1.5 times larger than the seeing limit
of the M33 data. Thus while it is possible that search radii
in the range of 10-20 pc represent some kind of match to the intrinsic
properties of OB associations, it is also possible that
the results using larger search radii are more dependent on the
underlying density of field stars,
and thus any search radius below some maximum value
would produce similar results in all galaxies.

For a more detailed comparison of the Galactic data with the
M33 and NGC 6822 data we will
fix the search radius at the value used in the M33 study,
22 pc (\markcite{w91}Wilson 1991). 
The OB associations identified by the algorithm using
a 22 pc search radius are shown in Figure~\ref{fig-6}. Of the
48 associations with 3 or more members, the algorithm 
(1) identifies 21 compact associations very well,
(2) identifies tight groups of stars in 10 of the associations but
misses the outlying members, (3) identifies two or more small groups
in  15 larger associations, and (4) completely fails to identify 2
associations. In comparison, the M33 data reveal 195 groups containing
at least 3 members, some of which are likely to be
chance superpositions of unrelated stars (see \S 3). The
median properties of the groups in the two galaxies are
very similar (diameter $\sim 20 $ pc, number of members $\sim 5$).
Of the 31 associations with 10 or more members, the
algorithm identifies tight groups of  at least
10 stars in 12 of the associations, while most of the remaining associations
are broken up into two or more groups with less than 10 stars.
These 12 groups have a median diameter of 60 pc and a median of
18 stars identified as members. These results compare quite
favorably to the associations identified with the same
physical search radius in M33 (median diameter 60 pc; median of 15 members)
and in NGC 6822 (median diameter 40 pc; median of 14 members; only 3
associations identified).
Subject to the discussion above, these 
results suggest that if we are interested in studying rich, 
compact OB associations, using a search radius of a fixed physical
size, and one that is not too large (about 20 pc), 
is the best approach to take in studying OB associations
in different galaxies (see also \markcite{w92}Wilson 1992).

\section{Conclusions}

We have used two Galactic sample of OB stars and associations to test 
an automatic grouping algorithm used to identify
extragalactic OB associations. By using a sample of stars for
which association membership is already known, we can test whether
the automatic algorithm correctly identifies true physical
OB associations. The main results are summarized below.

(1) The algorithm does a good job of identifying the known OB
associations using a search radius which is defined by the
stellar surface density, $R_s = \sqrt{1/\pi\Sigma} = 78$ pc. No known
associations are missed, although a few nearby
neighbors cannot be separated into two associations by the algorithm.
The sample of Galactic OB associations identified in previous
surveys thus appears to be a reasonably uniformly selected sample,
despite the many different techniques used to determine 
association membership.

(2) The presence of a handful of groups identified by the algorithm
but not known to be OB associations is used to estimate the level
of contamination by chance groupings of unrelated stars.
For the Galactic data,
the contamination by chance groupings is 
$\sim 45\%$ if groups as small as 3 members are included, and is 
$\sim 25\%$ if only groups containing 6 or more members are counted
as associations. 
This contamination is smaller than that obtained from
a random distribution of stars,
and suggests that truly random distributions of stars may provide
an overly pessimistic estimate of the contamination by random groups
obtained with this algorithm.

(3) The Galactic OB associations with at least
10 members identified using a search radius
of 78 pc have mean diameters of 150 pc, substantially larger
than associations identified using a similarly defined
search radius (22 pc) in the inner kiloparsec of M33. Applying the same
algorithm to the Galactic data 
with a search radius of 22 pc identifies only tight clumps of stars within
the larger OB associations. However, the sizes and
number of members of these tight clumps agree quite well
with the properties of OB associations identified in M33
and NGC 6822 using the same physical search radius.
This result
suggests that we should use the same
value for the search radius {\it in parsecs} to identify
the associations, rather than tying the search radius to
the local stellar surface density. This approach is most
easily justified for comparing two galaxies with similar molecular
cloud populations and star formation efficiencies, but is applicable under
many conditions to galaxies with different cloud populations
and star formation efficiencies.

(4) For Galactic associations with more than 10 members identified by the
algorithm, 25\% are in fact two or more nearby associations
which cannot be separated by the algorithm. Thus there is
some danger that the largest groups identified in nearby
galaxies may contain more than one OB associations, which
has implications for understanding the co-evality of star
formation in associations. 

C. D. W. would like to thank Roberta Humphreys for making available
machine-readable copies of the catalogs and for her referee comments
on the manuscript.
C. D. W. was partially supported by NSERC Canada through a Women's
Faculty Award and Research Grant. Part of this work was performed
while K. J. B. held an NSERC Targeted Female Undergraduate Summer Research
Award.

\clearpage

\figcaption[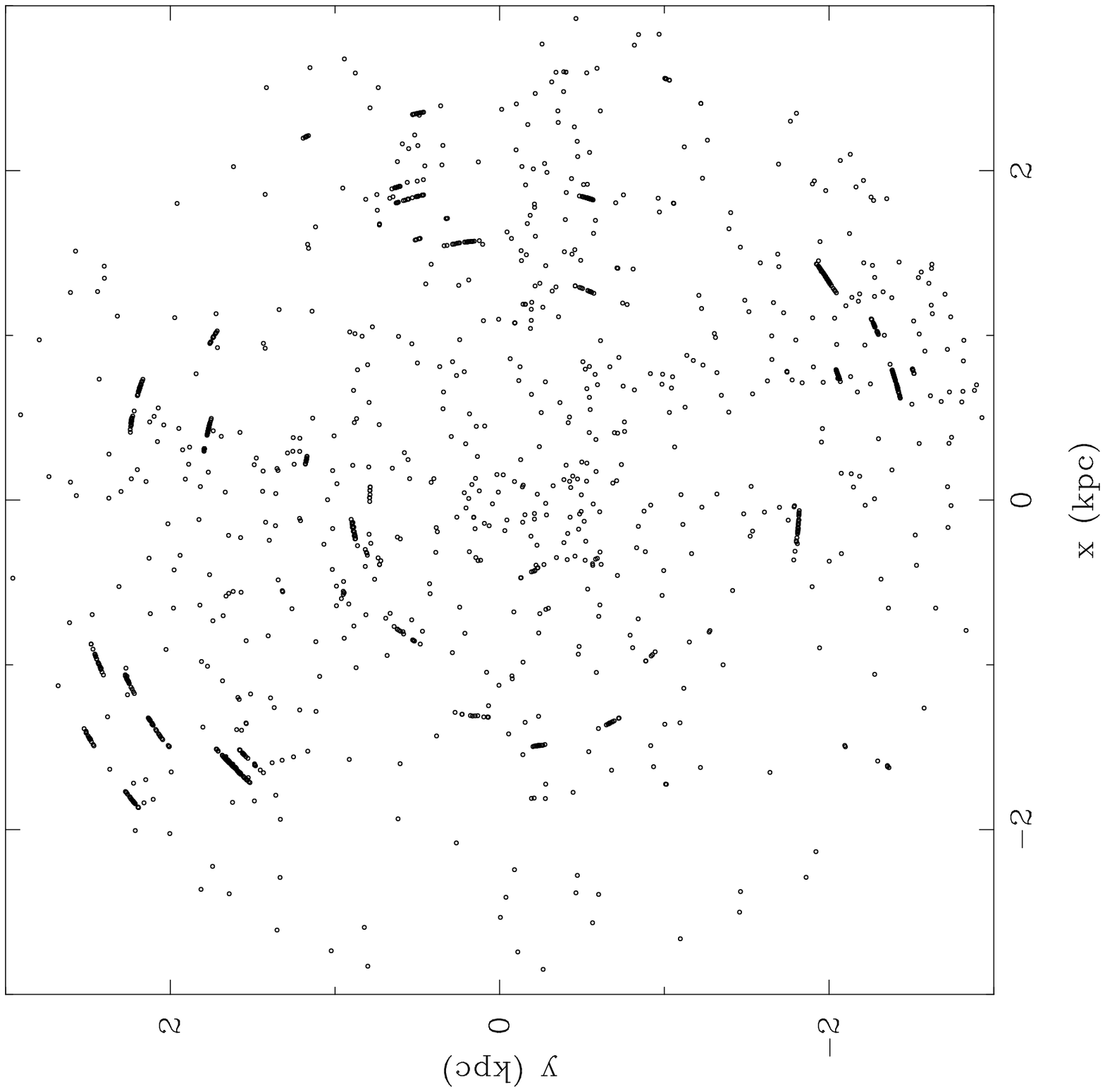]{(a) The distribution of OB stars brighter
than $M_v = -4.4$ mag lying within 3 kpc of the Sun from
the catalog of 
Humphreys
\& McElroy (1984) and 
Blaha \& Humphreys (1989). The short arcs
are the OB associations and the Galactic Center lies along the
positive X axis. (b) The same data, but with stars that are
members of OB associations smeared into a circularly
symmetric distribution (see text). \label{fig-1} }

\figcaption[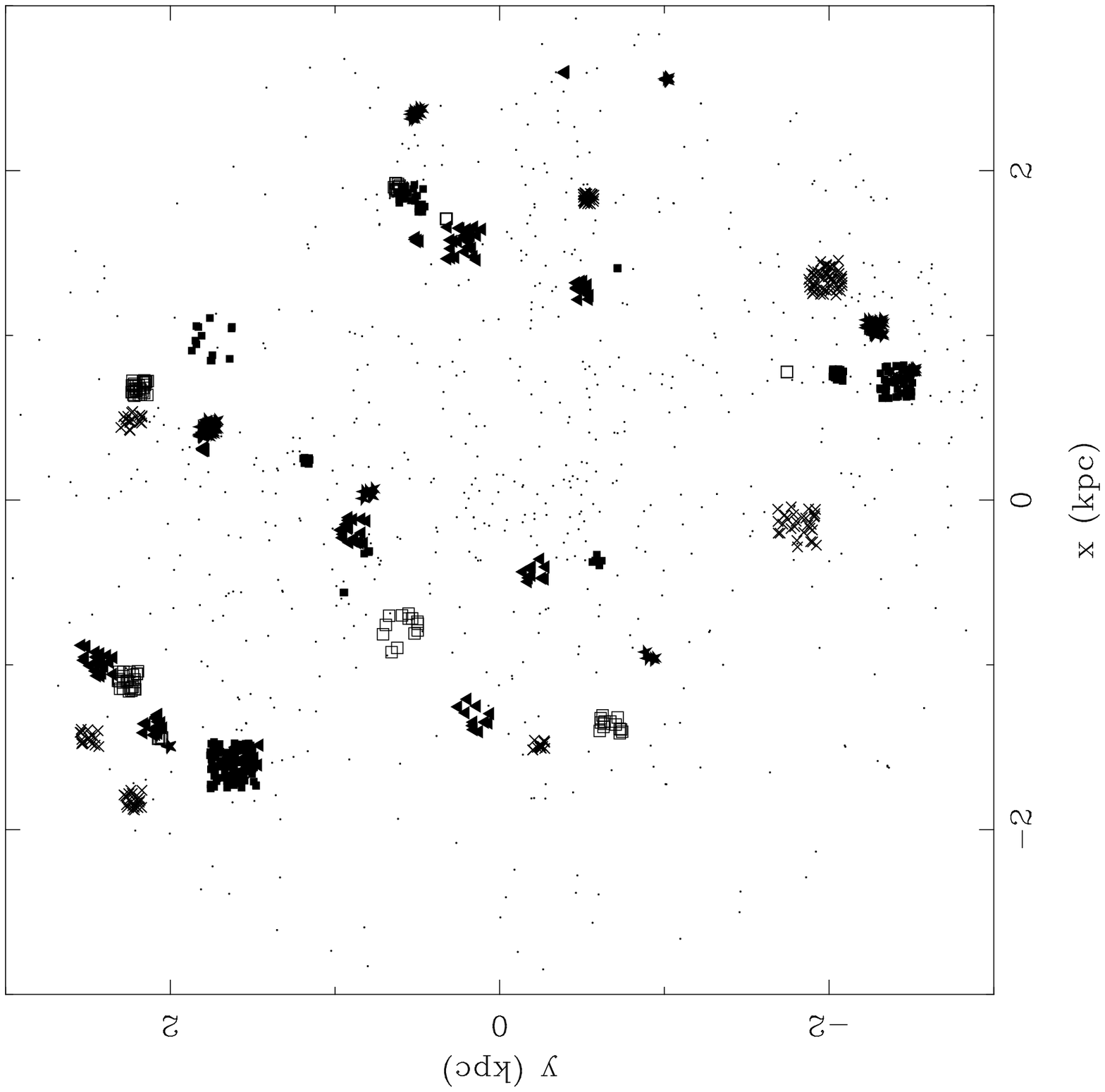,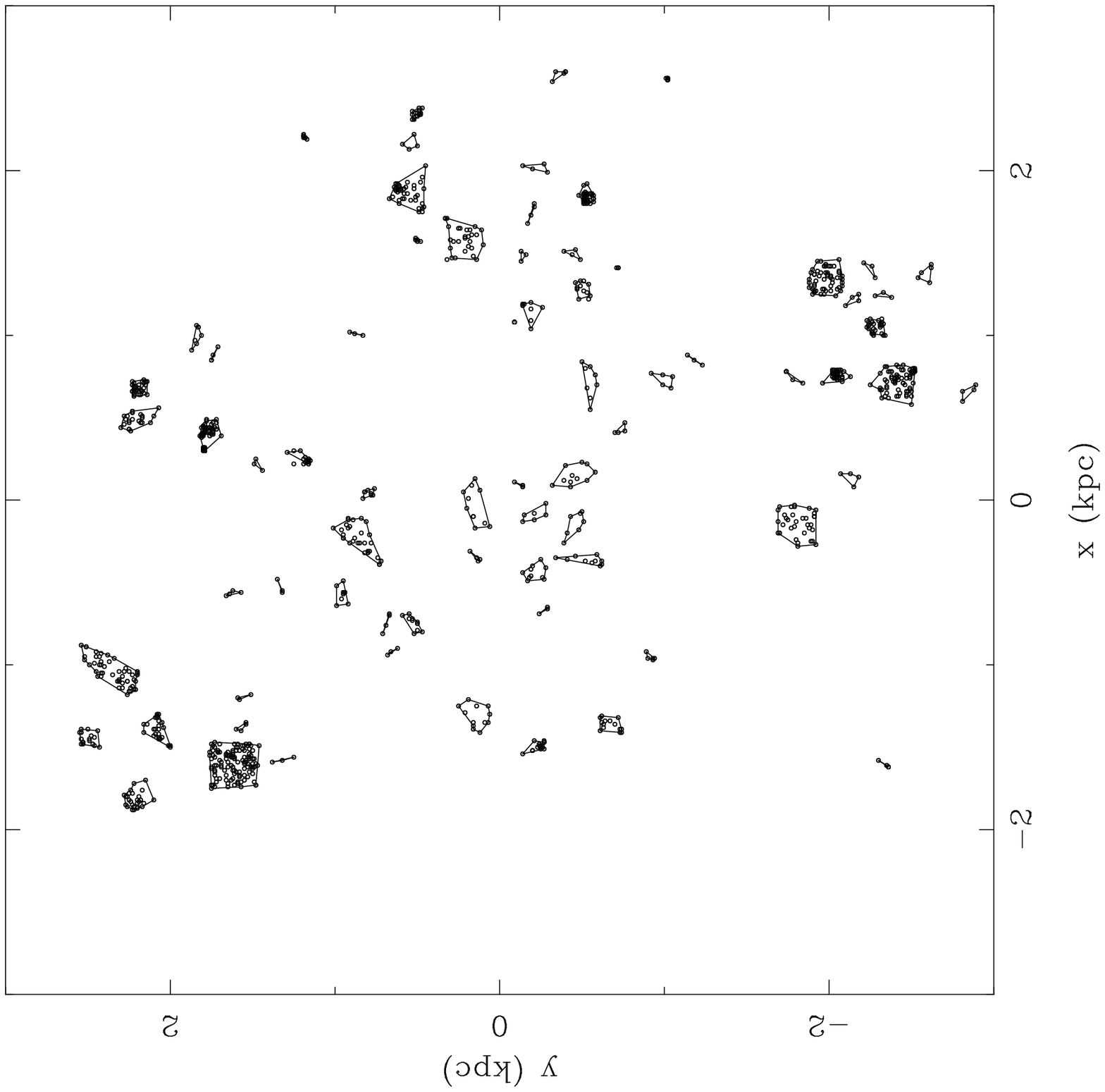]{(a) 
The distribution of OB associations and field
stars within 3 kpc of the Sun. Only stars brighter than
$M_v = -4.4$ are shown. The associations are identified
by the large symbols, the field stars by the small dots.
Only associations with at least 3 members brighter than
$M_v = -4.4$ are plotted with separate symbols. (b) The OB associations with
at least 3 members identified by the automated ``friends
of friends'' grouping algorithm. Stars outside the
associations are not shown. A search radius of 150 pc
was used to identify the OB associations. \label{fig-2} }

\figcaption[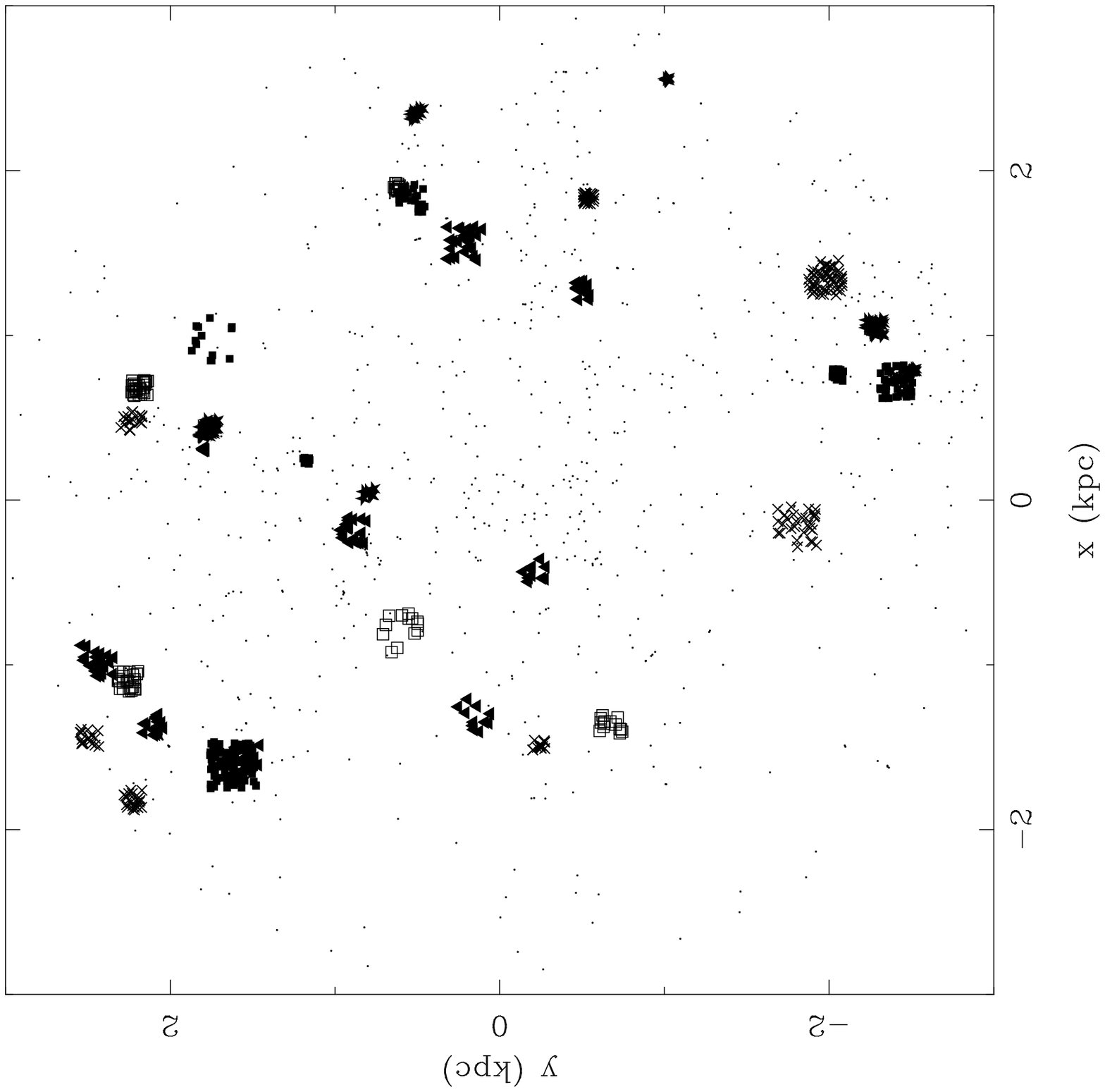,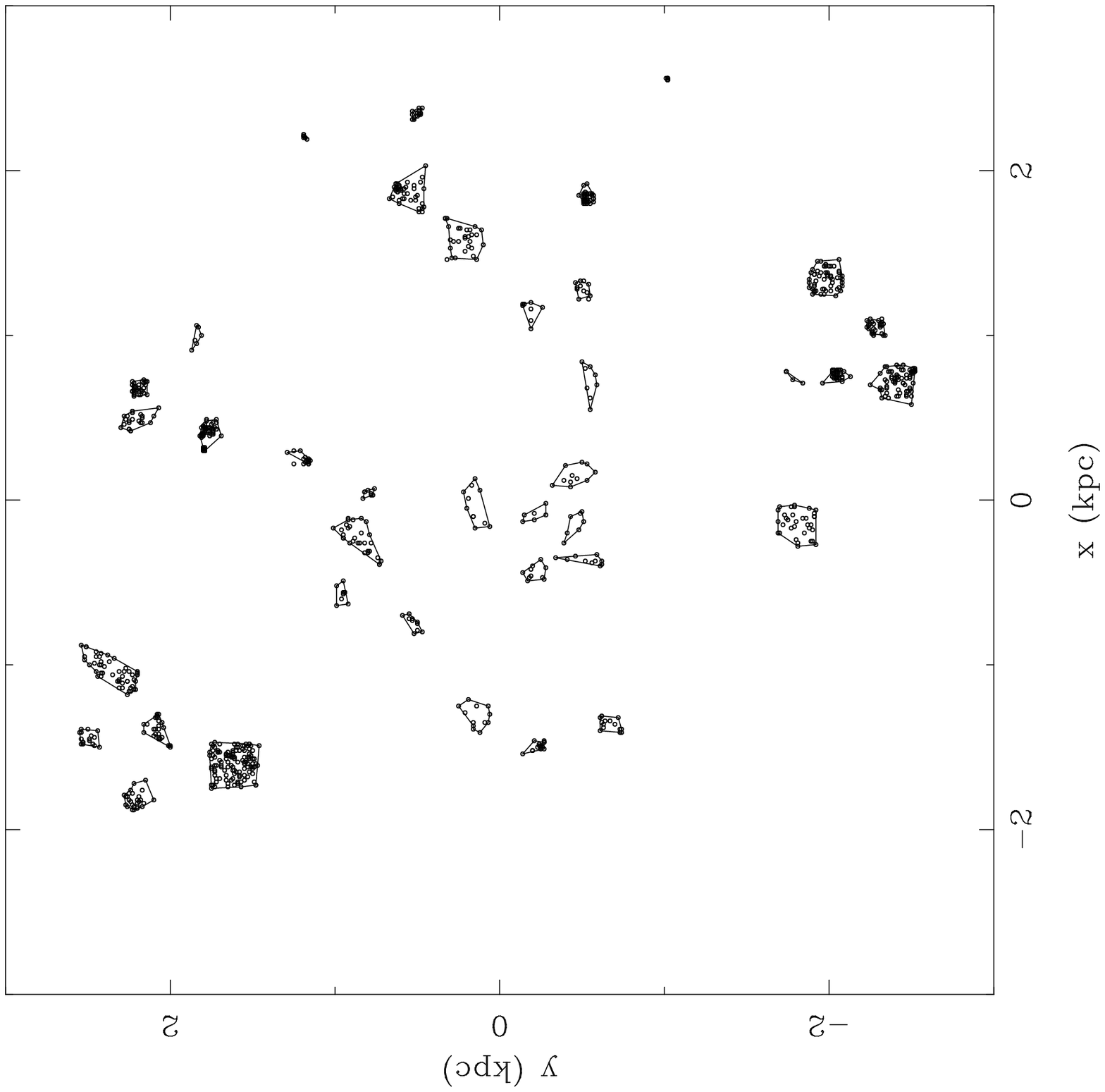]{The same as 
Figure 2,
but for associations
with at least 6 members. \label{fig-3} }

\figcaption[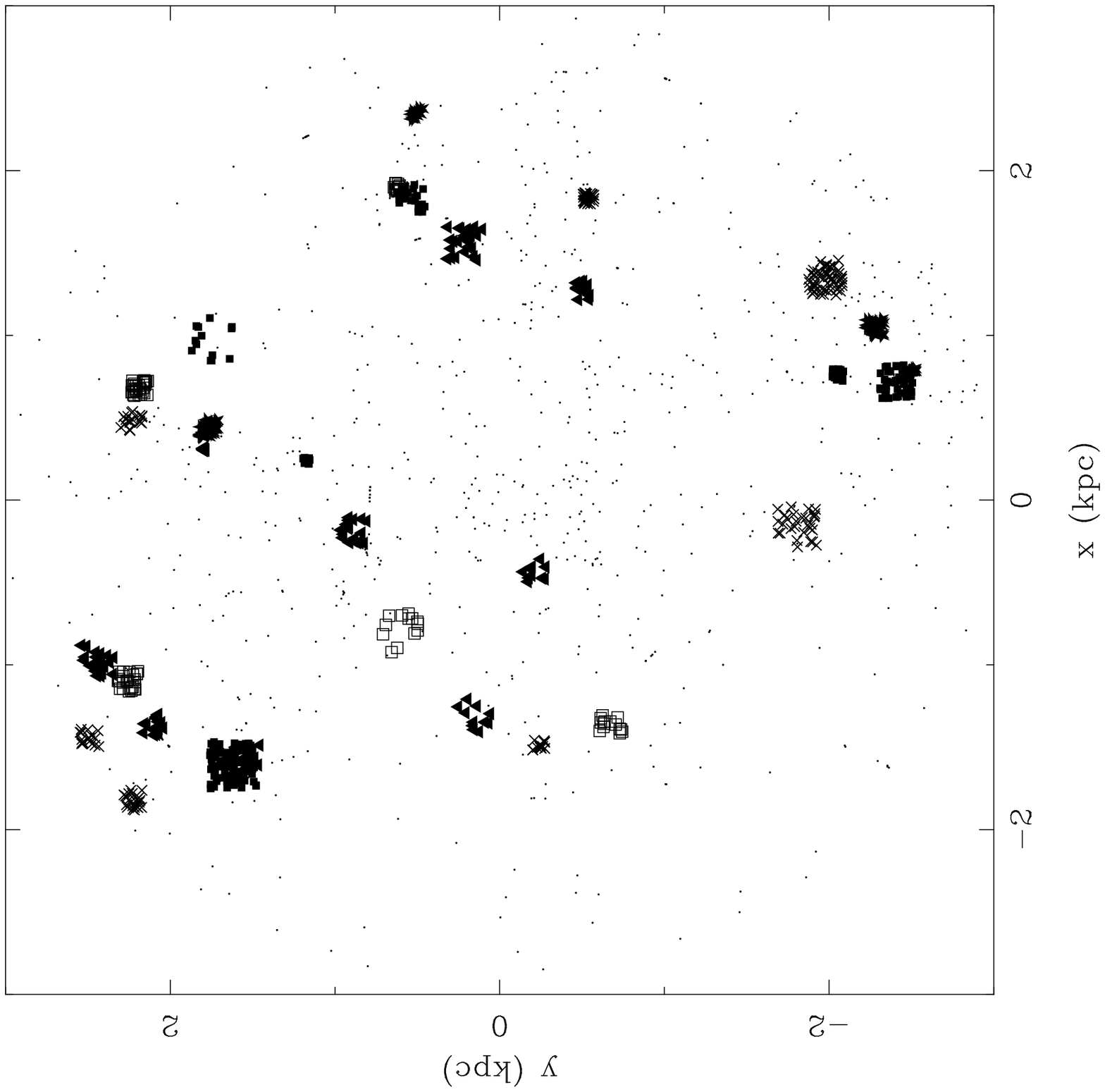,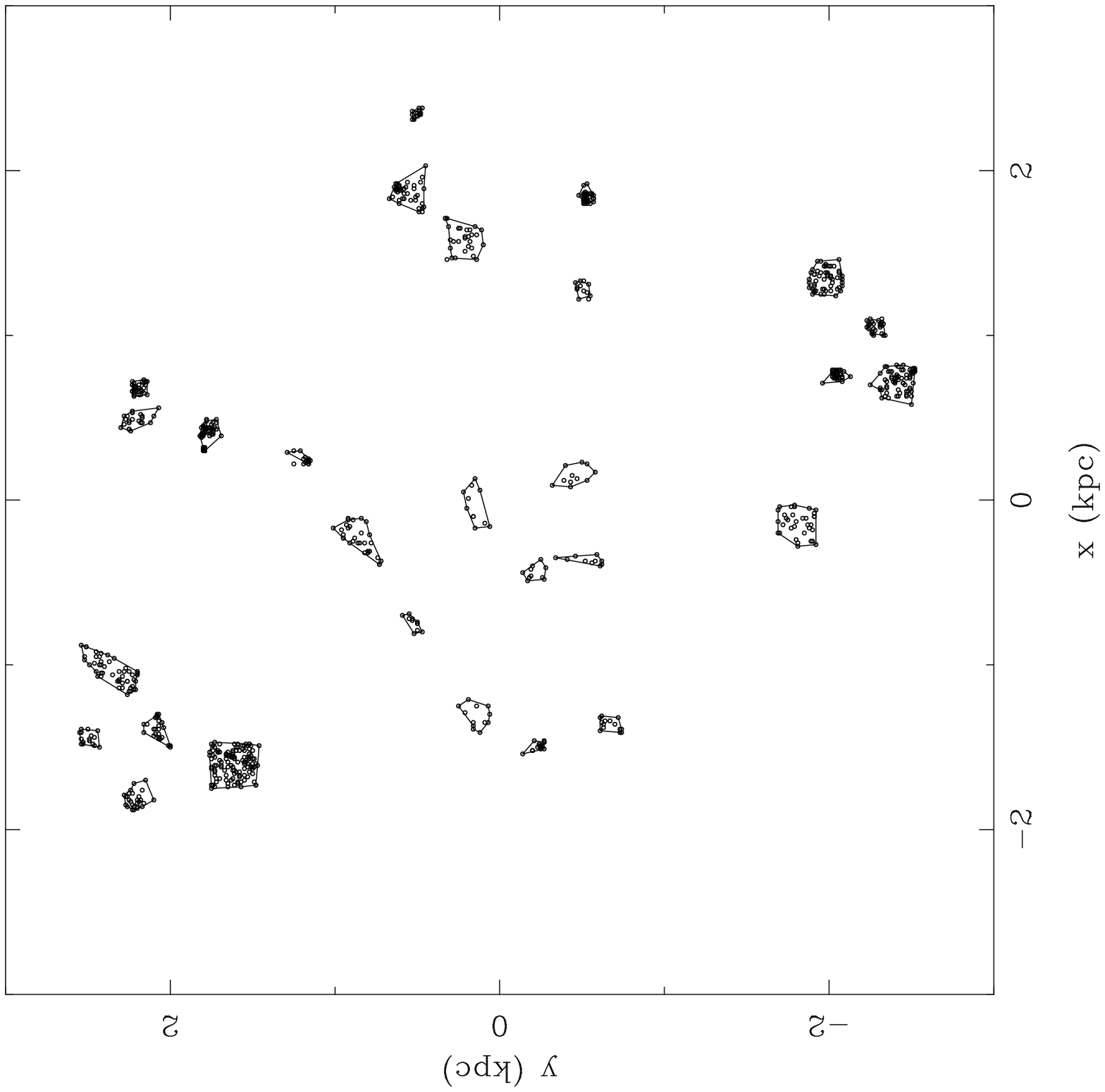]{The same as 
Figure 2,
but for associations
with at least 10 members. \label{fig-4} }

\figcaption[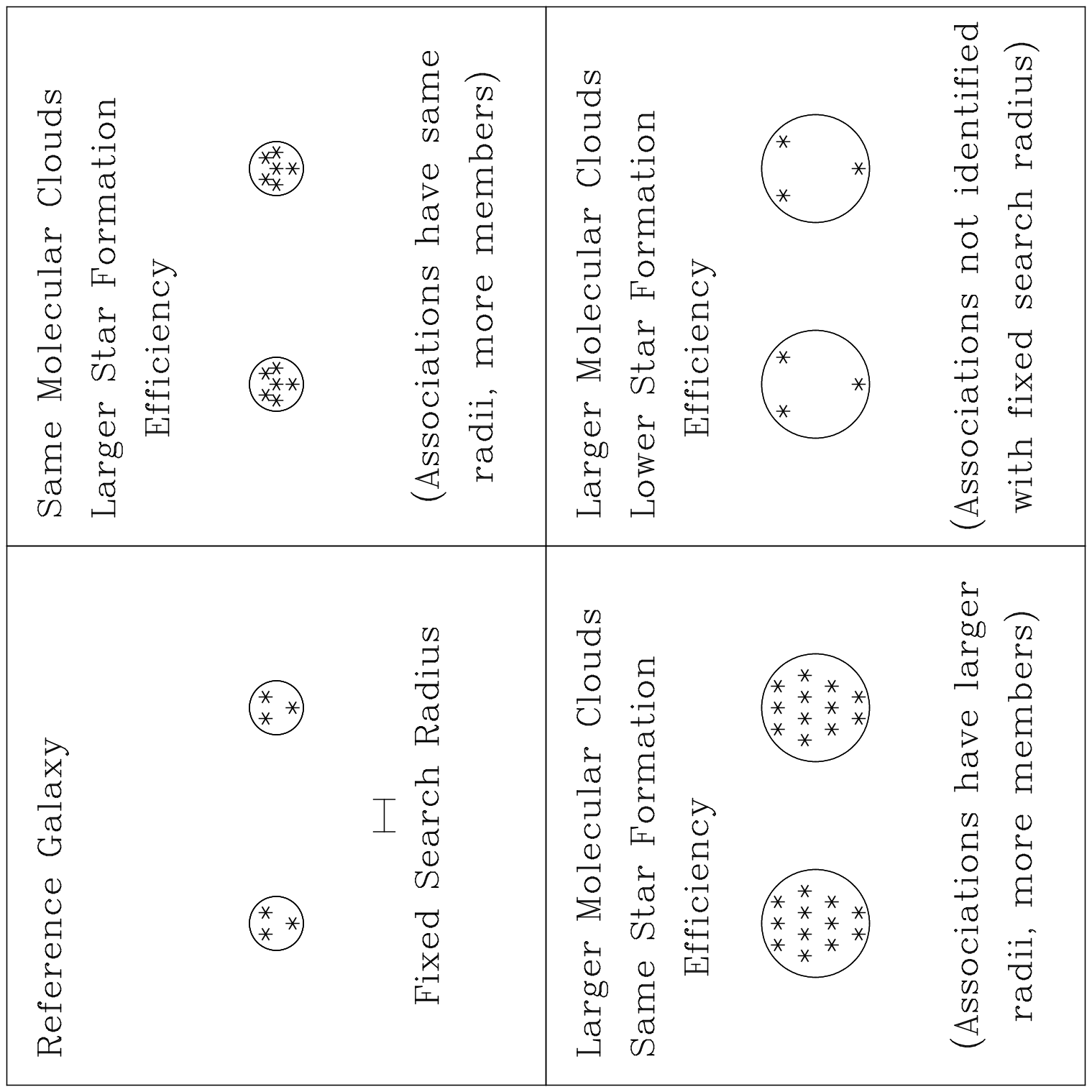]{Schematic illustration of how differences in the
underlying giant molecular cloud populations and star formation
efficiencies can be identified using a fixed physical search radius.
\label{fig-5}}

\figcaption[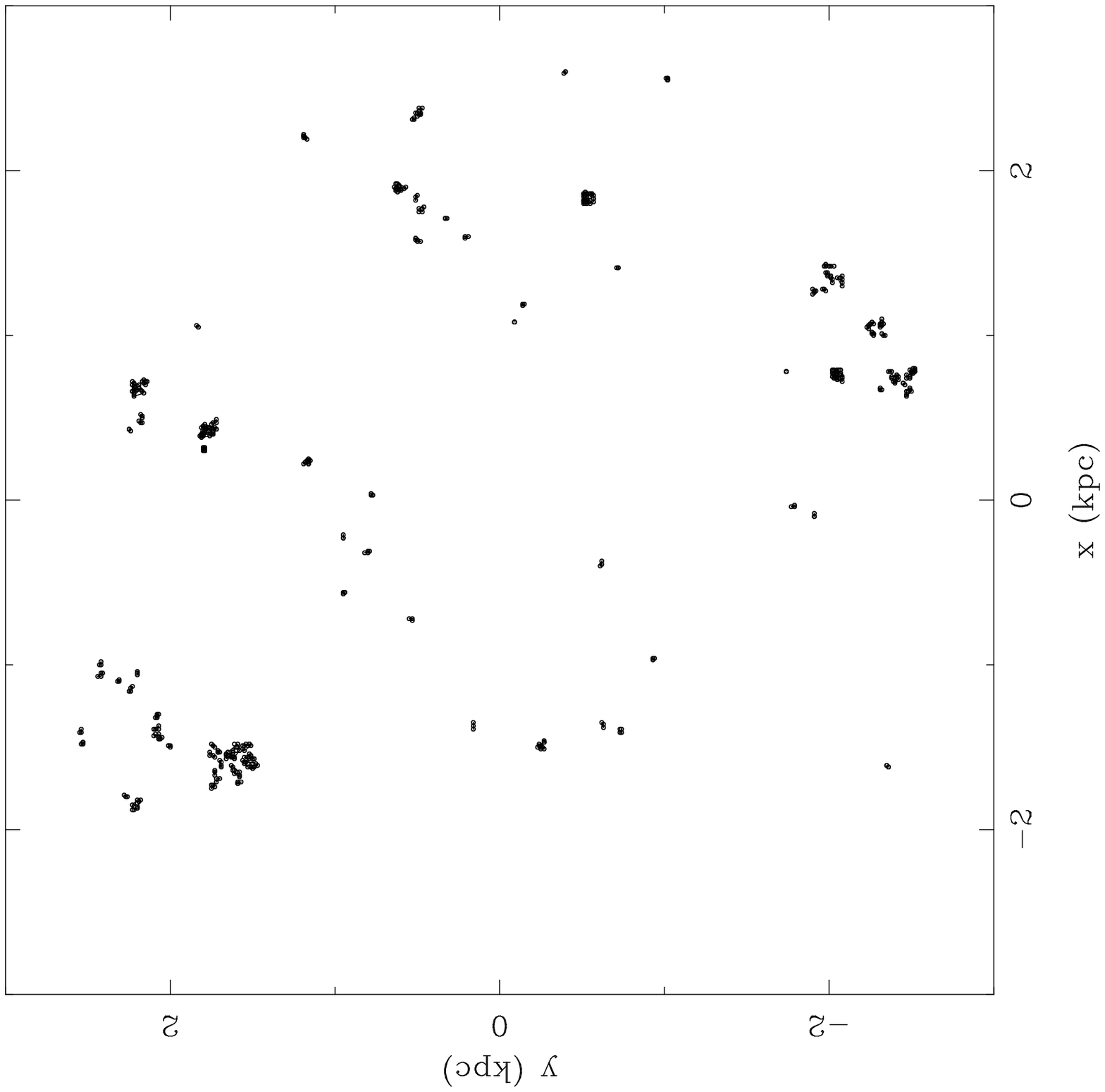]{The OB associations with
at least 3 members identified by the automated ``friends
of friends'' grouping algorithm using a search radius of 22 pc
to match the search radius used in studying OB
associations in M33. Compare with the distribution of OB associations and field
stars within 3 kpc of the Sun shown in 
Figure 2a.
\label{fig-6} }

\end{document}